# Bragg-Curve Simulation of Carbon-Ion Beams for Particle-therapy Applications: a study with the GEANT4 toolkit

M. Kh. Hamad

Physics Department, King Fahd University of Petroleum & Minerals, Dhahran, 31261, Saudi Arabia.

**Abstract**

We used the GEANT4 Monte Carlo MC Toolkit to simulate carbon ion beams incident on water, tissue, and bone, taking into account nuclear fragmentation reactions. Upon increasing the energy of the primary beam, the position of the Bragg-Peak transfers to a location deeper inside the phantom. For different materials, the peak is located at a shallower depth along the beam direction and becomes sharper with increasing electron density NZ. Subsequently, the generated depth dose of the Bragg curve is then benchmarked with experimental data from GSI in Germany. The results exhibit a reasonable correlation with GSI experimental data with an accuracy of between 0.02 and 0.08 cm, thus establishing the basis to adopt MC in heavy-ion treatment planning. The Kolmogorov-Smirnov K-S test further ascertained from a statistical point of view that the simulation data matched the experimentally measured data very well. The two-dimensional isodose contours at the entrance were compared to those around the peak position and in the tail region beyond the peak, showing that bone produces more dose, in comparison to both water and tissue, due to secondary doses. In the water, the results show that the maximum energy deposited per fragment is mainly attributed to secondary carbon ions, followed by secondary boron and beryllium. Furthermore, the number of protons produced is the highest, thus making the maximum contribution to the total dose deposition in the tail region. Finally, the associated spectra of neutrons and photons were analyzed. The mean neutron energy value was found to be 16.29 MeV, and 1.03 MeV for the secondary gamma. However, the neutron dose was found to be negligible as compared to the total dose due to their longer range.

## 1- Introduction

In recent years, particle therapy with protons and heavier ions like [12]C has begun to elicit increasing interest in radiation treatment [1-4]. The main advantage of using ions over photons is primarily attributed to a favorable profile of the dose depth of ions, which commences as a flat low-dose region, and then increases in depth until the Bragg Peak [5-6], which can be located at the targeted region during treatment planning [7]. Compared to protons, on the other hand, there are some advantages of using heavier ions like carbon [8-14]. Firstly, and due to the mass difference, these ions exhibit fewer multiple scatterings than protons in lateral directions. As a result, there is less range of straggling as beams penetrate deep inside the body to the final treatment locations. This ensures that sharper field edges can be achieved, an important consideration for tumors near to critical structures. Furthermore, the energy loss, based on the basic Bethe-Bloch formula, is proportional to the squared projectile charge $z_p^2$, but not its mass [15]. Consequently, the energy loss of a carbon ion is 36 times greater than that of a proton of the same velocity, which therefore implies that localized biological damage is much higher, with higher cell killing and fewer chances for repair [16]. Secondly, the ionization rate increases at the end of the particle range resulting in clusters of lesions on the DNA molecule in the cells [17]. Given that DNA lesion clusters are much more difficult to repair than one DNA damage, carbon ions have increased biological efficacy [18]. In contrast to the above advantages, once the ions penetrate matter with energies in the MeV range, they are partially fragmented [1]. The fragments have a longer range and a wider distribution of energy than the primary ions, producing a distinctive dose tail beyond the peak [18]. Since this feature could affect healthy tissues, it needs to be considered in any comprehensive treatment planning [19]. One can estimate this effect if the residual nuclides of carbon ions in relevant biological materials like tissue and bone are known. However, this carbon ion fragmentation has still not been fully investigated. Therefore, more measurements and experimental knowledge are needed for a better understanding of the production of nuclear fragments in biological matter. In a previous study [20], the distribution of the fluence and the yield of the nuclear fragments were calculated by the FLUKA code. In addition, the dose distribution from the [12]C ion is evaluated by the FLUKA code [1]. The neutrons from the fragmentation of light nuclei in water using Geant4 MC were also investigated by Pshenichnov *et al* [21]. In the current study, we use the GEANT4 MC (v9.3.2) [22-23] to investigate the Bragg-Peak for a delivered dose of carbon-12 ([12]C) ion beams incident on water, tissue, and bone, taking into account nuclear fragmentation reactions. The resulting depth dose Bragg curve for the $^{12}$C ion beam in water is then compared to the experimental data from the GSI [7, 21]. Finally, the data were analyzed using the Kolmogorov-Smirnov K-S test [25] for the 270 MeV/u $^{12}$C beam in water to assess how well the simulation represents the experimental data from a statistical point of view.

## 2- Materials and methods

Geant4 is a toolkit for the simulation of the passage of particles through matter based on the Monte Carlo method. The hadron therapy setup was carefully modeled. The passive beamline was completely simulated. Besides, the used materials have been validated using a special Geant4 code to be sure that they are of the correct compositions and densities. This was achieved by calculating the attenuation coefficient ($\tau$) using $I = I_o e^{\frac{-\mu}{\rho}\rho\tau}$, and then compare the results with the corresponding data obtained from the XCOM database. The phantom, which is a box of uniform material filled with water tissue or bone, contains identical voxels where data on the dose deposited by primary and secondary particles can be collected. The use of the water phantom is needed by the international protocol on the measure of the dose in the case of proton and ion beams [24]. Fig. 1 illustrates a carbon ion beam that perpendicularly hits a 30x4x4 cm$^3$ detector inside a 40x40x40 cm$^3$ water phantom, where the distance between the phantom and the beam source is 20 cm in air. The position and size of the detector are changeable to match the requirements of the measurement.

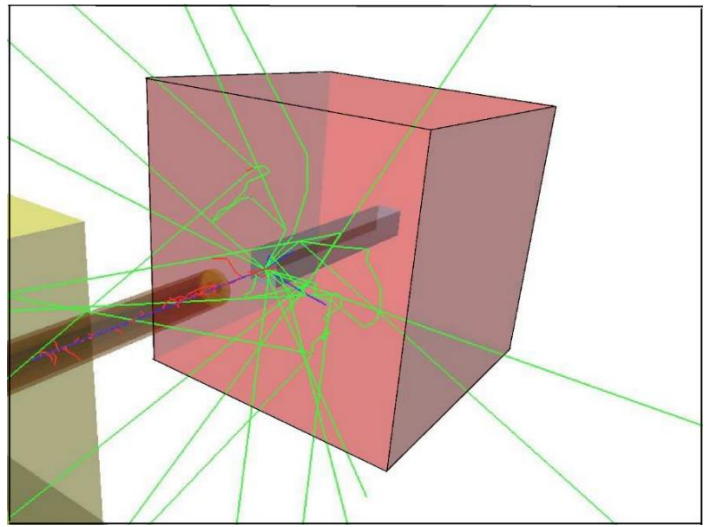

*Figure 1: A carbon ion beam (blue) in air hits a movable and variable size detector (grey) inside a phantom box (pink). The secondaries shown are negative charged particles (red), positive charged particles (blue), and neutral particles (green).*

Two different techniques for the setup of detectors were adopted; the first one was assigned to measure the energy distribution and the dose depth profile in one dimension (Bragg curve). As required by the incident particle energy, a 30x4x4 cm$^3$ detector was used for this purpose. It contains 3000 slabs with each one having a 0.01x4x4 cm$^3$ volume. The second setup aims at obtaining the two-dimensional isodose contours perpendicular to the beam direction around the Bragg Peak positions and in the fragment tail region. Each isodose plot covers a 1x4x4 cm$^3$ volume representing a one cm thick slab in the beam direction. The slab is divided laterally into 1600 identical voxels with a volume of 1.0x0.1x0.1 cm$^3$ for each.

In the simulation, the tracking of particles continues down to a threshold, below which no secondary will be generated, and consequently the particle energy is locally deposited. This threshold is defined as a range cut-off, which is internally converted to an energy for individual materials. The cut-off value is critical because it should be small relative to the voxel dimensions, but large enough not to cause very slow simulation runs. We chose the cut-off in our simulation, after careful investigation, to be 0.05 mm.

### 3- Results and discussions

A test for the matching between the experimental data from the GSI [7, 21] and the simulated Bragg Peaks for $^{12}$C at different energies (135, 195, 270, and 330 MeV/u) in water has been conducted. The Bragg-Peak positions for $^{12}$C at different energies are shown in Table 1. Generally, as the energy of the primary ion increases, the peak position is transferred to the deeper parts of the target material and the absorbed dose around the peak position declines. Besides, the accuracy of the simulation was between 0.02 and 0.08 cm. Fig. 2 (a) shows the experimental data for the 270 MeV/u $^{12}$C ion beam (red circles) which has been interpolated to form a set of data (dashed red line) that has a common abscissa (depth) with the simulated data (blue line). The 270 MeV/u $^{12}$C beam in different media is shown in Fig. 2 (b); the Bragg-peak position was 14.42 cm inside the water phantom and 14.15 cm inside the tissue, while it was at 8.51 cm in the bone phantom (Fig. 2 b). As a result, the peak is located at a shallower depth (stronger stopping power) along the beam direction (X-axis) and becomes sharper in the medium with a higher electron density NZ (Table 2), which is in line with the literature (see [36]).

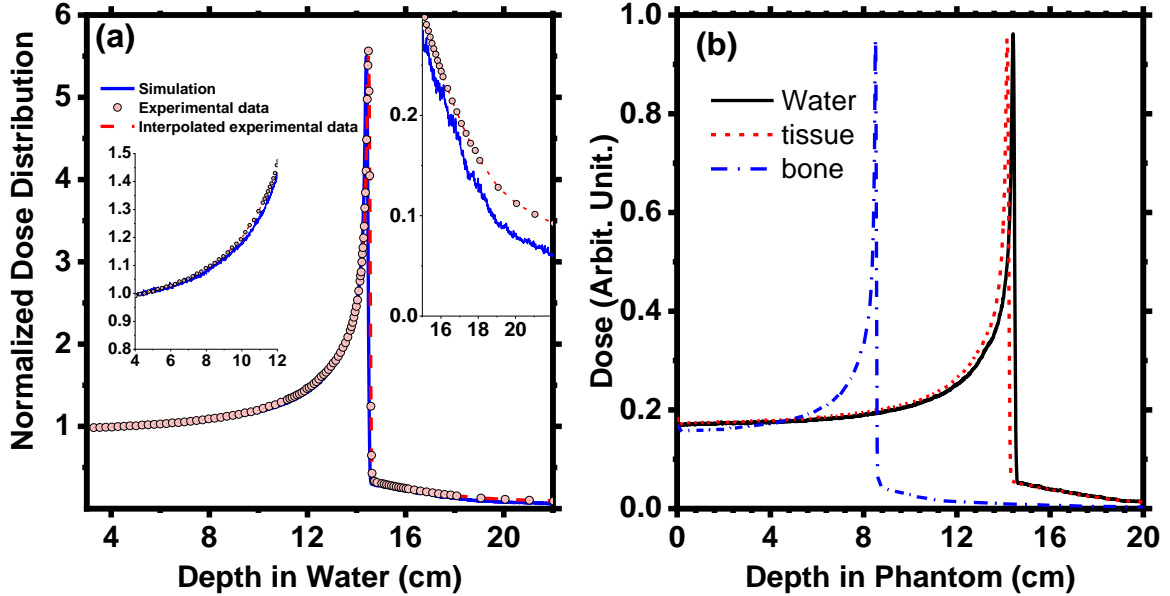

*Figure 2: (a) Simulated vs. experimental depth dose distributions for 270 MeV/u $^{12}$C beam in water. The original experimental data (red circles) has been interpolated to form the red dashed line. The simulation (black line) was found in excellent agreement with the experimental results except in the tail region beyond the Bragg Peak (see text). (b) Bragg Peaks of 270 MeV/u $^{12}$C beam in water, tissue and bone.*

*Table 1: Simulated and experimental Bragg-Peak positions for $^{12}$C ions at different incident energies.*

| Energy (MeV/u) | Simulation Peak (cm) | Experimental Peak (cm) | Difference (%) |
| --- | --- | --- | --- |
| 135 | 4.36 | 4.34 | 0.46 |
| 195 | 8.28 | 8.34 | 0.72 |
| 270 | 14.42 | 14.49 | 0.48 |
| 330 | 20.09 | 20.13 | 0.20 |

*Table 2: Simulated Bragg-Peak positions and the FWHM for 270 MeV/u $^{12}$C beams in three different media.*

| Medium | Peak position (cm) | FWHM (cm) |
|---|---|---|
| Water | 14.42 | 0.419 |
| Tissue | 14.15 | 0.373 |
| Bone | 8.51 | 0.203 |

To assess how well the simulation represents the experimental data from a statistical point of view, we perform the Kolmogorov-Smirnov *K-S* test [25] for the 270 MeV/u $^{12}$C beam in water. The test covers the range between 4 and 19 cm from the entrance point, thus covering the entrance, the Bragg Peak itself, and the region in the tail beyond the peak in which most of the fragments deposit their energy (secondary doses). It is worth mentioning that beyond 19 cm, the deposited energy becomes very small, and the simulated curve starts to deviate from the experimental data (compare the right and left insets of Fig. 2a). We have comprehensively investigated this behavior by lowering the energy cut-off to very low values. The small deviation can therefore only be explained by ambient background readings in the experiment that are not included in the simulation, and that only affect the comparison when the beam-related energy deposition approaches zero. We run the *K-S* statistical test using the *R* software environment for statistical computing and graphics [26]. The concept of the *K-S* test is to evaluate the Cumulative Distribution Fraction CDF function of the two data sets and calculate the maximum vertical deviation *D* between both functions. A small deviation *D* (*i.e.,* the two cumulative distributions of the data set are not very different) implies that the simulation data match the experimentally measured data very well. More details about the K-S test are available in, for example, [27-31]. Fig. 3 shows where both *CDF* functions coincide completely. The results of the *K-S* test were $D = 0.053$, with a corresponding *p*-value that is equal to 0.984. This *p*-value, which is close to one, implies that the two data sets are mutually consistent. This result identifies a crucial part of the code validation and provides the basis for the further investigation reported in this research with reliability.

For completeness, we constructed isodose contours in planes perpendicular to the beam direction. The contours are calculated in slices around and beyond the Bragg Peak position for the three studied phantoms. To establish a reference for the dose distribution, Fig. 4 shows the isodose in the YZ plane for a one cm slice perpendicular to the beam and located exactly at the entrance of the water phantom. At the beam entrance to the phantom, the total absorbed dose is mainly due to the primary ions. As the beam penetrates the phantom material, more and more secondaries are produced, part of which have a longer range compared to the beam. The total absorbed dose around the peak position is then a mixture between both the primary and secondary doses (fragments). Fig. 5(a) shows the isodose in the three investigated phantom materials around the peak position. Due to the reduced lateral straggling associated with the shorter projected range in bone, the corresponding Bragg Peak is narrower than that in water and tissue (Fig. 2b and Table 2), and this is reflected as a less laterally distributed dose (Fig. 5a). In addition, it is important to focus on the differences between energy deposition trends in water and tissue. Usually, the patient is represented by a water phantom, and experimental data is acquired in water, while tissue is defaulted as water. This might be true for high energy x-rays, but for particle radiation any differences in $Z$ could lead to differences in the dose and the location of the Bragg Peak.

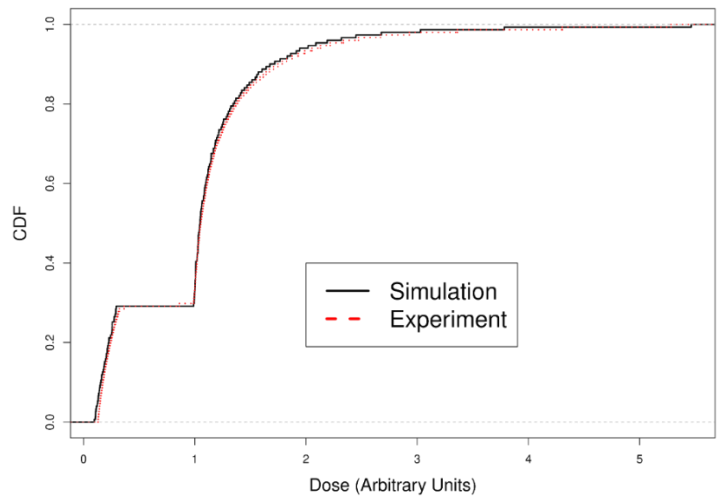

*Figure 3: The cumulative distribution functions CDF of the simulated Bragg Peak and of the experimental data. Both CDF functions coincide, and therefore the maximum vertical deviation D was found very small (see text).*

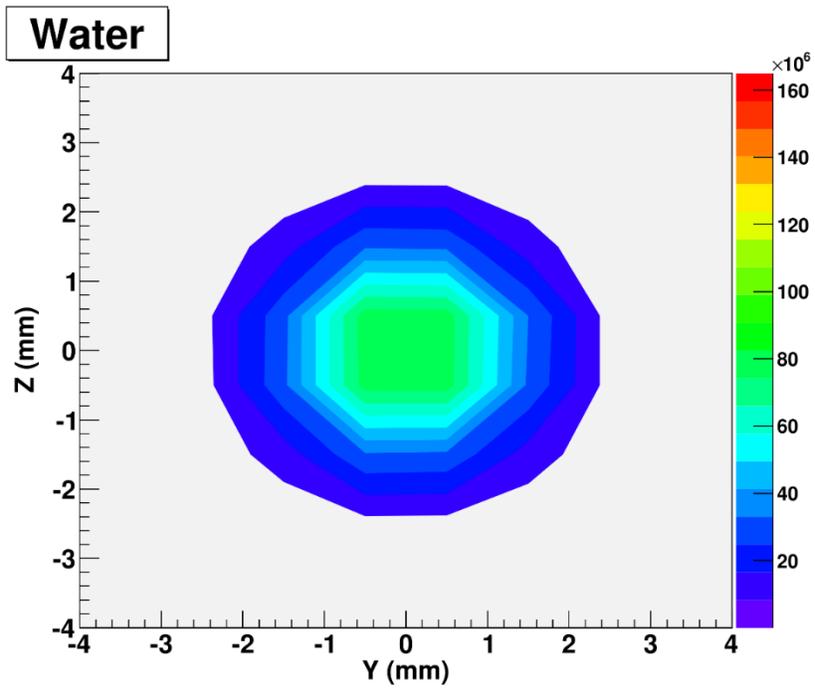

*Figure 4: Reference isodose for a 270 MeV/u $^{12}$C beam calculated in one cm slice at the entrance of the water phantom.*

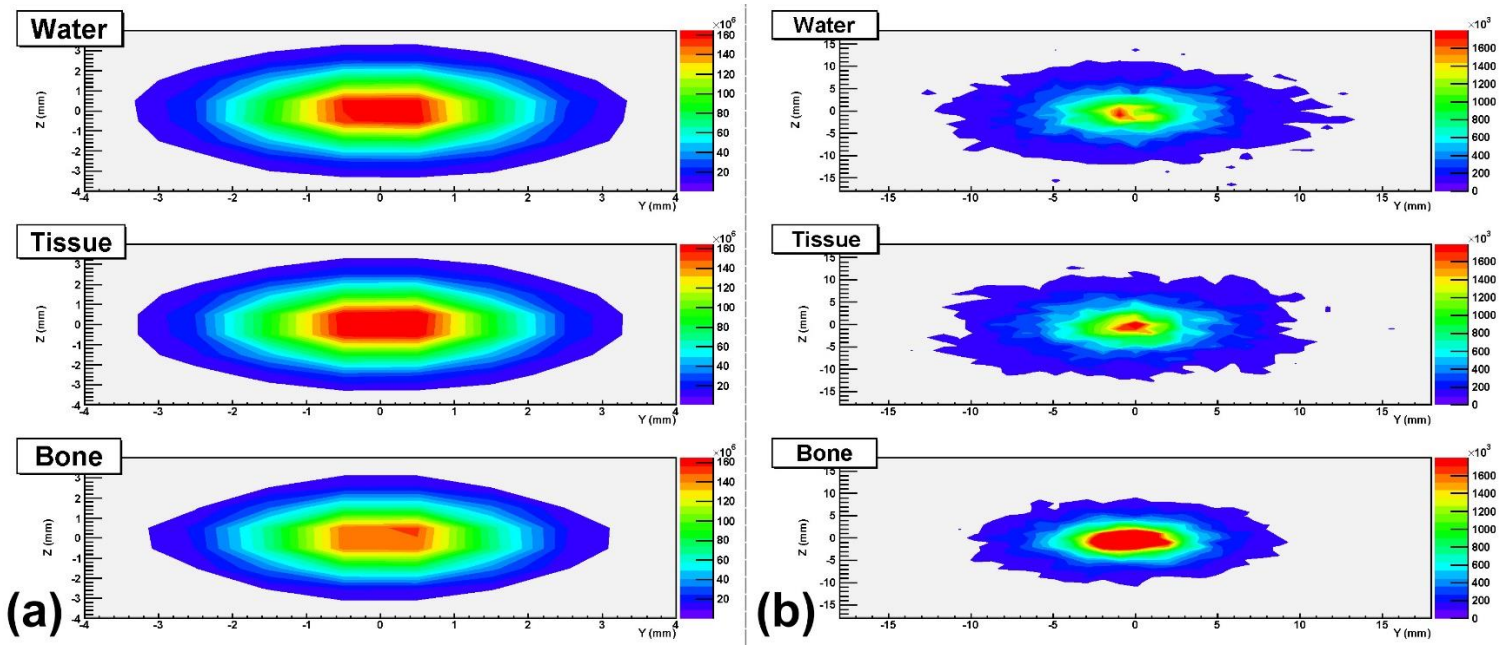

*Figure 5: Two-dimensional isodose from 270 MeV/u $^{12}$C beam in the three phantom materials, (a) around and (b) beyond the Bragg Peak position.*

On the other hand, the fragmentation dose contribution obviously appears to be dominant beyond the peak position after the primary beam has completely been stopped. Fig. 5(b) shows the isodose in the three materials. The relatively stronger fragmentation in bone is obvious in this figure because the dose is purely due to the fragments, while this effect was immersed in the much stronger primary beam before and at the Bragg Peak (Fig. 5a). Finally, comparing the isodose at and beyond the Bragg Peak (Fig. 5(a) and (b), respectively) to the reference isodose at the entrance (Fig. 4), provides a qualitative illustration of the physics of angle straggling and fragmentation in the three investigated media.

### *3.1 Secondary doses analysis*

To acquire a comprehensive calculation of the dose, nuclear fragmentation reactions should be thoroughly investigated. According to Hultqvist *et al* [32], exposure to secondary radiation during particle therapy is of great concern because of possible tissue damage and the risk of secondary cancer induction. The dose deposition beyond the peak is completely due to the nuclear fragmentation reactions. Fragments ranging from Z=1 (hydrogen) to Z=9 (fluorine) were identified, with a minority of even heavier particles (Fig.6 (a) and (b)). The number of protons and oxygen ions obviously dominate (Fig. 6(a)), a finding consistent with the fact that the phantom under study consists of water. Because of their longer ranges, protons and alpha particles are expected to be responsible for most of the dose deposition beyond the Bragg Peak. The dose from heavier fragments is, on the other hand, effectively merged in the direct beam energy deposition at and below the stopping point of the primary carbon nuclei. On the other hand, boron and secondary carbon ions delivered the highest energy deposited per fragment (Fig. 6(b)), followed by beryllium and nitrogen. However, the number of these fragments is much smaller than the number of protons and oxygen ions knocked out in the water phantom. The large number of protons compared to beryllium, for example (see Fig. 6(a)), is reflected in Fig. 6(b) as a dense array of points, though the energy per proton is in general less than that of beryllium.

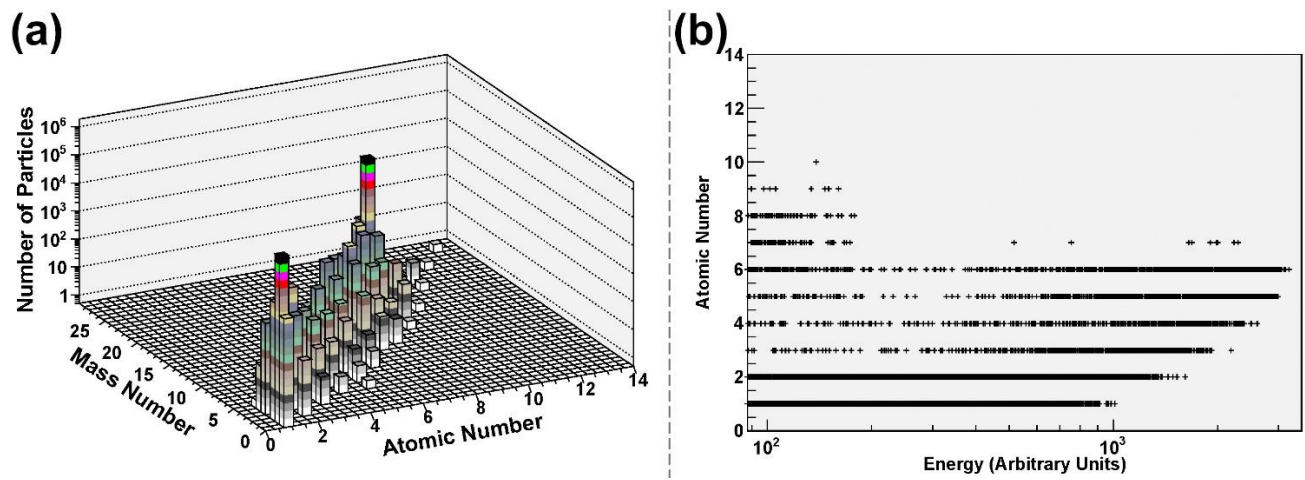

Figure 6: (a) Relative number of fragments, characterized by their atomic and mass number, produced beyond the Bragg Peak in the water phantom by 270 MeV/u $^{12}C$ beam. (b) Energy per secondary ion, produced in nuclear interactions of 270 MeV/u $^{12}C$ primaries in water.

The fragmentation reactions discussed so far are only considered to be charged particles. In order to comprehend the analysis, neutrons and gamma particles that do contribute to the dose should also be investigated. Little information is known about the dose from secondary neutrons created in heavy-ion irradiations [21]. Fast neutrons produced in projectile nuclei fragmentation provide extra doses inside and outside the phantom. The mean free path of these fast neutrons in matter is commonly much longer compared with the ranges of ions of the same energy. Therefore, the neutron-associated dosage is predicted to be distributed in a much larger volume compared to protons and heavier ions. The energy distribution due to secondary neutrons from a 270 MeV/u $^{12}C$ beam is presented in Fig. 7(a). Though the spectrum covers a broad energy range, the mean neutron energy value was found to be 16.29 MeV. Similar behavior is reported by both Pshenichnov *et al* [21], and Soltani-Nabipour *et al* [1]. On the other hand, gamma photons can be emitted in different ways during the passage of ions in the phantom; the mass difference in nuclear reactions can appear as photons (direct reaction process). In addition, if the reaction reaches an excited state of a fragment, the transitions between energy levels will result in gamma radiation. The annihilation of positrons from positron emitting fragments also contributes to the photon yield inside and outside the phantom. Fig. 7 (b) presents the energy distribution for secondary gamma photons. Again, the mean energy was 1.03 MeV, although the tail of the spectrum reaches much higher energy values. These doses potentially cause secondary cancers and induce other harmful effects [32]. Recently, a neutron detector was designed, and located behind the target under carbon irradiations, to measure the induced neutrons by Khorshidi *et al* [33-34]. They claimed that the fragmentation of projectile nuclei and the local energy deposition of charged hadrons are inevitable consequences for dose absorption. As a result, the shielding of the heavy-ion medical accelerator is essential [35]

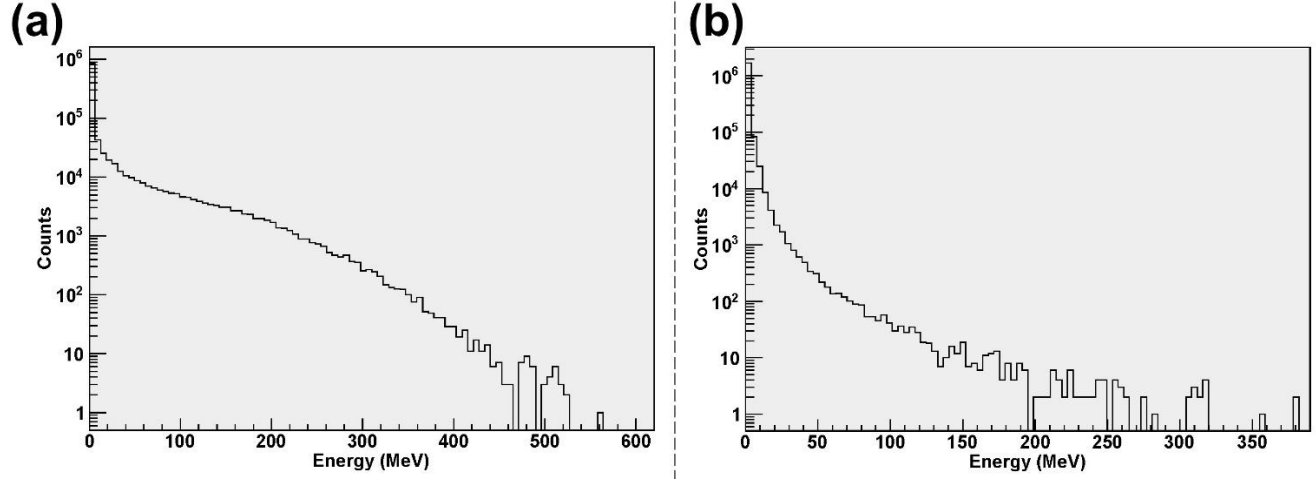

*Figure 7: Energy distribution due to secondary (a) neutrons and (b) gamma from 270 MeV/u $^{12}$C beam in water.*

To shed more light on the fragmentation reactions, we shot 62 MeV/u $^{12}$C ions (*i.e.,* 744 MeV) onto the water phantom. The results were carefully analyzed in detail for selected events. Firstly, in event number 0, the primary $^{12}$C ion, which is located at a depth (in *mm*) of (x=1.66, y=-0.78, and z= 0.25), left with an energy equal to 668.91 MeV (~90% of primary energy), and underwent a nuclear reaction with $^{16}$O:

$$^{12}C + {}^{16}O \rightarrow {}^{11}C + {}^{15}N + p + n + 2\gamma \qquad (1)$$

This reaction gives subsequent secondaries tracks as shown in Table 3. The total kinetic energy of the products is about 631.22 MeV (~85% of primary energy), and the total gamma energy is about 6.71 MeV (~0.9% of the primary energy). Since the *Q*-value for this reaction is negative, part of the incident primary energy is required to overcome the associated threshold energy.

*Table 3: Tracking information for secondaries that have a parent ID = 1 from one $^{12}C$ ion at 744 MeV.*

| TrackID | Product | K.E(MeV) | Process |
|---------|---------|----------|---------|
| 867 | $^{11}C$ | 539.46 | IonInelastic |
| 866 | γ | 5.14 | IonInelastic |
| 865 | $^{15}N$ | 42.78 | IonInelastic |
| 864 | γ | 1.57 | IonInelastic |
| 863 | n | 14.38 | IonInelastic |
| 862 | p | 34.6 | IonInelastic |

It is worth mentioning that this collision produced an unstable $^{11}C$ isotope (see Eq. 1) with a half-life time of 20.3 minutes, and emitted a positron (β+):

$$^{11}C \rightarrow {}^{11}B + e^+ + \nu \qquad (2)$$

The annihilation of the positron produces two gamma-rays (each with 0.511 keV) that can be detected from outside the phantom with appropriate detectors. This provides one way to visualize the range of the original carbon beam in the patient. Thus, this methodology is suitable for positron-emission-tomography (PET). Moreover, the neutron with an energy of 14.38 MeV (Table 3) obtained from the first primary interaction happens to produce a proton, electron, and an anti-neutrino by (β-) decay. The neutron has a track ID = 863. As a result, for parent ID = 863, a proton with an energy of 2.81 MeV inside the phantom was recognized. Since neutrons have long ranges due to their neutral nature, most neutrons will interact outside the phantom in the physical treatment room. Secondly, the primary carbon ions lost their energy by means of multiple scattering and ionization down to rest (events number 1 up to 5). At each step, electrons were ejected with some energy and were tracked. As a matter of fact, these electrons travel beyond the voxel of interaction and contribute to the extended lateral and longitudinal dose profiles. Finally, when a primary carbon interacts with hydrogen in water (the phantom), the result is an ion inelastic scattering process (event number 6):

$$^{12}C + {}^{1}H \rightarrow p + {}^{12}C + 2\gamma \qquad (3)$$

The proton may, for example, travel beyond the Bragg Peak position. At an even higher beam energy, the nuclear processes become more complicated. For example, 270 MeV/u (3240 MeV)

$^{12}$C ions undergo inelastic processes with hydrogen, (Table 4), to give three alpha particles and a proton:

$$^{12}C + {}^{1}H \rightarrow 3\,\alpha + p \tag{4}$$

Table 4: Tracking information for secondaries that have a parent ID = 1 from one $^{12}$C at 3240 MeV.

| Track ID | Products | Energy (MeV) | Process |
|---|---|---|---|
| 275 | $^4$He | 991.508 | IonInelastic |
| 274 | $^4$He | 999.234 | IonInelastic |
| 273 | $^4$He | 1178.27 | IonInelastic |
| 272 | $^1$H | 29.2358 | IonInelastic |

The highly energetic $^4$He particle that has a track ID = 274 interacts with $^{16}$O:

$$^{4}He + {}^{16}O \rightarrow {}^{15}O + {}^{3}H + p + n + \gamma \tag{5}$$

More noticeable in this reaction is the $^{15}$O (track ID = 11467) which is also a positron emitter with a half-life of 122 seconds. This positron will annihilate with an electron to two detectable 511 keV photons, which can be used for PET. These results reveal that heavy ion beams from elements that span the periodic table can be tracked using PET methods to ensure the beam's effectiveness in tumor irradiation.

### 4- Conclusions

In the present study, results of the Monte Carlo MC Hadron therapy model available in the GEANT4 toolkit exhibit reasonable correlation with the GSI experimental data, hence establishing the basis for the adoption of MC in heavy ion treatment planning. In the tail region, the dose deposition is completely attributable to nuclear fragmentation reactions that produce fragments ranging from hydrogen to mostly fluorine, with some minorities above Z=9. The maximum energy deposited per fragment was mainly generated by secondary carbon ions, followed by secondary boron and beryllium. Nevertheless, the total number of protons released was the largest, therefore contributing the most to the total dose deposition in the tail region. The effect of the phantom material has also been studied, showing that bone generates more dose due to nuclear fragmentation compared with both water and tissue. The distributions of the gamma and the neutron energy were measured, and further analysis of the expected interactions outside the phantom is still to be reported.

**Acknowledgment**

The author would like to acknowledge the support provided by the Deanship of Research at King Fahd University of Petroleum & Minerals (KFUPM), Saudi Arabia.

Special thanks are due to Dr. Dieter Schardt (GSI) for providing us with the tables of experimental data on depth – dose distributions. Great gratitude is extended to Prof. Saed Dababneh who is no longer with us. His valuable advices and kind notices are deeply thanked.

**REFERENCES**

[1] J. Soltani-Nabipour, A. Khorshidi, F. Shojai, K. Khorami, Evaluation of dose distribution from 12C ion in radiation therapy by FLUKA code, Nuclear Engineering And Technology. 52 (2020) 2410-2414. doi:10.1016/j.net.2020.03.010.

[2] O. Jakel, Medical physics aspects of particle therapy, Radiation Protection Dosimetry. 137 (2009) 156-166. doi:10.1093/rpd/ncp192.

[3] O. Jäkel, D. Schulz-Ertner, C. Karger, A. Nikoghosyan, J. Debus, Heavy Ion Therapy: Status and Perspectives, Technology In Cancer Research & Treatment. 2 (2003) 377-387. doi:10.1177/153303460300200503.

[4] D. Schulz-Ertner, O. Jäkel, W. Schlegel, Radiation Therapy With Charged Particles, Seminars In Radiation Oncology. 16 (2006) 249-259. doi:10.1016/j.semradonc.2006.04.008.

[5] W. Bragg, and R. Kleeman. On the α particles of radium, and their loss of range in passing through various atoms and molecules, The London, Edinburgh, And Dublin Philosophical Magazine And Journal Of Science. 10 (1905) 318-340. doi:10.1080/14786440509463378.

[6] J.S. Nabipour, A. Khorshidi, Spectroscopy and optimizing semiconductor detector data under X and g photons using image processing technique, J. Med. Imag. Radiat. Sci. 49 (2) (2018) 194e200, https://doi.org/10.1016/j.jmir.2018.01.004.

[7] L. Sihver, D. Schardt, T. Kanai, Depth-Dose Distributions of High-Energy Carbon, Oxygen and Neon Beams in Water, Japanese Journal Of Applied Physics. 18 (1998). doi: https://doi.org/10.11323/jjmp1992.18.1_1.

[8] M. Hultqvist, J. Lillhök, L. Lindborg, I. Gudowska, H. Nikjoo, Nanodosimetry in a 12C ion beam using Monte Carlo simulations, Radiation Measurements. 45 (2010) 1238-1241. doi:10.1016/j.radmeas.2010.05.033.

[9] G. Kraft, Tumor therapy with heavy charged particles, Progress In Particle And Nuclear Physics. 45 (2000) S473-S544. doi:10.1016/s0146-6410(00)00112-5.

[10] G. Kraft, M. Scholz, U. Bechthold, Tumor therapy and track structure, Radiation And Environmental Biophysics. 38 (1999) 229-237. doi:10.1007/s004110050163.

[11] S. Brons, G. Taucher-Scholz, M. Scholz, G. Kraft, A track structure model for simulation of strand breaks in plasmid DNA after heavy ion irradiation, Radiation And Environmental Biophysics. 42 (2003) 63-72. doi:10.1007/s00411-003-0184-9.

[12] P. Azimi, A. Movafeghi, Proton therapy in neurosurgery: a historical review and future perspective based on currently available new generation systems, Int. J. Clin. Neurosci. 3 (2) (2016) 59e80, https://doi.org/10.22037/icnj.v3i2.13324

[13] S. Malmir, A. Asghar Mowlavi, S. Mohammadi, Enhancement evaluation of energy deposition and secondary particle production in gold nanoparticle aided tumor using proton therapy, Int. J. Canc. Manag. 10 (10) (2017), e10719, https://doi.org/10.5812/ijcm.10719.

[14] A. Khorshidi, Accelerator-based methods in radio-material 99Mo/99mTc production alternatives by Monte Carlo method: the scientific-expedient considerations in nuclear medicine, J. Multiscale Model. (JMM) 11 (1) (2020) 1930001, https://doi.org/10.1142/S1756973719300016.

[15] E. Segrè, H. Staub, H. Bethe, et al, Experimental nuclear physics, 1st ed., John Wiley & Sons, New York, 1953.

[16] Alonso, Jose R. Review of ion beam therapy: Present and Future. United States: N. p., Web. https://www.osti.gov/servlets/purl/765471. (2000).

[17] Hall, E.J. Radiobiology for the Radiologist. Fourth Edition, Lippincott Williams & Wilkins. (1993).

[18] E. Haettner, Experimental study on carbon ion fragmentation in water using GSI therapy beams., MSc, KTH Royal Institute of Technology in Stockholm, 2006.

[19] P. Petti, A. Lennox, Hadronic Radiotherapy, Annual Review Of Nuclear And Particle Science. 44 (1994) 155-197. doi:10.1146/annurev.ns.44.120194.001103.

[20] J. Soltani-Nabipour, M. Popovici, G. Cata-Danil, Residual Nuclei Produced by 290 MeV/u $^{12}$C ions beam in a liquid water target., Romanian Reports In Physics. 62 (2010) 37-46. http://194.102.58.21/2010_62_01/art04Soltanidoc.pdf (accessed 9 September 2020).

[21] I. Pshenichnov, I. Mishustin, W. Greiner, Neutrons from fragmentation of light nuclei in tissue-like media: a study with the GEANT4 toolkit, Physics In Medicine And Biology. 50 (2005) 5493-5507. doi:10.1088/0031-9155/50/23/005.

[22] Geant4 User's Guide (2012). https://geant4.web.cern.ch/support/user_documentation

[23] A. Lechner, V. Ivanchenko, J. Knobloch, Validation of recent Geant4 physics models for application in carbon ion therapy, Nuclear Instruments And Methods In Physics Research Section B: Beam Interactions With Materials And Atoms. 268 (2010) 2343-2354. doi:10.1016/j.nimb.2010.04.008.

[24] Dosimetry and Medical Radiation Physics Section, Absorbed Dose Determination in External Beam Radiotherapy, The International Atomic Energy Agency, Vienna, 2000. http://www-pub.iaea.org/mtcd/publications/pdf/trs398_scr.pdf (accessed 9 September 2020).

[25] S. Dababneh, E. Al-Nemri, J. Sharaf, Application of Geant4 in routine close geometry gamma spectroscopy for environmental samples, Journal Of Environmental Radioactivity. 134 (2014) 27-34. doi:10.1016/j.jenvrad.2014.02.019.

[26] R Development Core Team, R: A Language and Environment for Statistical Computing R Foundation for Statistical Computing, Vienna, Austria. http://www.R-project.org

[27] A. Cuadra-Sánchez, J. Aracil, Finding the Optimal Aggregation Period, Traffic Anomaly Detection. (2015) 11-27. doi:10.1016/b978-1-78548-012-6.50002-3.

[28] E. Mocanu, P. Nguyen, M. Gibescu, Deep Learning for Power System Data Analysis, Big Data Application In Power Systems. (2018) 125-158. doi:10.1016/b978-0-12-811968-6.00007-3.

[29] G. Christodoulakis, S. Satchel, The validity of credit risk model validation methods, The Analytics Of Risk Model Validation. (2008) 27-43. doi:10.1016/b978-075068158-2.50006-8.

[30] R. Woods, Validation of Registration Accuracy, Handbook Of Medical Image Processing And Analysis. (2009) 569-575. doi:10.1016/b978-012373904-9.50043-x.

[31] J. Blackledge, Statistical Modelling and Analysis, Digital Image Processing. (2005) 512-540. doi:10.1533/9780857099464.4.512.

[32] M. Hultqvist, I. Gudowska, Secondary doses delivered to an anthropomorphic male phantom under prostate irradiation with proton and carbon ion beams, Radiation Measurements. 45 (2010) 1410-1413. doi:10.1016/j.radmeas.2010.05.020.

[33] A. Khorshidi, Neutron activator design for 99Mo production yield estimation via lead and water moderators in transmutation's analysis, Instrum. Exp. Tech. 61 (2) (2018) 198-204, https://doi.org/10.1134/S002044121802015X.

[34] A. Khorshidi, Molybdenum-99 production via lead and bismuth moderators and milli-structure-98Mo samples by the indirect production technique using the Monte Carlo method, Phys. Usp. 62 (9) (2019) 931-940, https://doi.org/10.3367/UFNe.2018.09.038441.

[35] M. Ashoor, A. Khorshidi, L. Sarkhosh, Appraisal of new density coefficient on integrated-nanoparticles concrete in nuclear protection, Kerntechnik 85 (1) (2020) 9-14, https://doi.org/10.3139/124.190016.

[36] Knoll, G., 1999. Radiation Detection And Measurement. 3rd ed. New York: John Wiley & Sons, pp.31-32.